\documentclass[aps,prl,twocolumn,groupedaddress]{revtex4}
\usepackage{graphicx}
\usepackage{amsmath,amssymb,amsfonts}
\usepackage{bm}

\begin{document}


\title{Quantum and Tunnelling Capacitance in Charge and Spin Qubits}


\author{R. Mizuta}

\affiliation{Hitachi Cambridge Laboratory, J. J. Thomson Ave., Cambridge, CB3 0HE, United Kingdom}
\author{R.~M.~Otxoa}
\affiliation{Hitachi Cambridge Laboratory, J. J. Thomson Ave., Cambridge, CB3 0HE, United Kingdom}
\author{A.~C. Betz}
\affiliation{Hitachi Cambridge Laboratory, J. J. Thomson Ave., Cambridge, CB3 0HE, United Kingdom}
\author{M. F. Gonzalez-Zalba}
\affiliation{Hitachi Cambridge Laboratory, J. J. Thomson Ave., Cambridge, CB3 0HE, United Kingdom}
\email[]{mg507@cam.ac.uk}

\date{\today}

\begin{abstract}

We present a theoretical analysis of the capacitance of a double quantum dot in the charge and spin qubit configurations probed at high-frequencies. We find that, in general, the total capacitance of the system consists of two state-dependent terms: The quantum capacitance arising from adiabatic charge motion and the tunnelling capacitance that appears when repopulation occurs at a rate comparable or faster than the probing frequency. The analysis of the capacitance lineshape as a function of externally controllable variables offers a way to characterize the qubits' charge and spin state as well as relevant system parameters such as charge and spin relaxation times, tunnel coupling, electron temperature and electron g-factor. Overall, our analysis provides a formalism to understand dispersive qubit-resonator interactions which can be applied to high-sensitivity and non-invasive quantum-state readout.

\end{abstract}


\maketitle

\section{I. INTRODUCTION}

In-situ state-readout of a quantum system via its dispersive interaction with a resonator is a promising technique that offers a compact and sensitive alternative to external mesoscopic detectors~\cite{Wallraff2004,Petersson2012, Colless2013, Gonzalez-Zalba2015,Viennot2015}. When applied to quantum information processing, dispersive phenomena can be exploited to obtain detailed information of the charge or spin qubit final state~\cite{Petersson2010}: The qubit's state-dependent capacitance causes a shift in the resonator frequency that can be readily detected by standard homodyne detection techniques~\cite{Schoelkopf1998,Angus2008, Barthel2009}. Therefore, understanding the origin of the qubit's capacitance at high frequencies becomes important for quantum-state readout~\cite{Chorley2012, Ciccarelli2011,Cottet2011,Schroer2012}. So far, different components of this capacitance have been identified: The quantum capacitance arising from adiabatic charge transitions and the non-zero curvature of the energy bands~\cite{Sillanpaa2005,Duty2005,Persson2010,Betz2015} and the tunnelling capacitance that appears when population redistribution processes, such as relaxation and thermal or resonant excitation, occur at a rate comparable or faster than the probing frequency~\cite{Ashoori1992,Persson2010a,Gonzalez-Zalba2015, Gonzalez-Zalba2016}.

In this letter, we present a description of the qubit's total capacitance that gathers both phenomena, quantum and tunnelling capacitance, under the same theoretical framework. We take the example of double quantum dots (DQD), an ideal system to host charge~\cite{Kim2015} and long-lived spin qubits~\cite{Veldhorst2015}. We describe the total capacitance of the system as a function of external variables such as energy detuning, temperature and magnetic field. We also identify two relevant timescales with regards to the operation of the resonator and the relaxation time of the system. Our results provide experimental tools to characterize the qubit charge and spin state as well as relevant qubit parameters such as tunnel coupling, electron temperature, electron g-factor and relaxation times. 

Finally, we note that our formalism is general to quantum multi-level systems and readily applicable to other types of qubits such as superconducting qubits~\cite{Pashkin2009} and hybrid qubits~\cite{Koh2012}.

\section{II. Differential Capacitance in Double Quantum Dots}

\begin{figure}[b]
\centering
\includegraphics{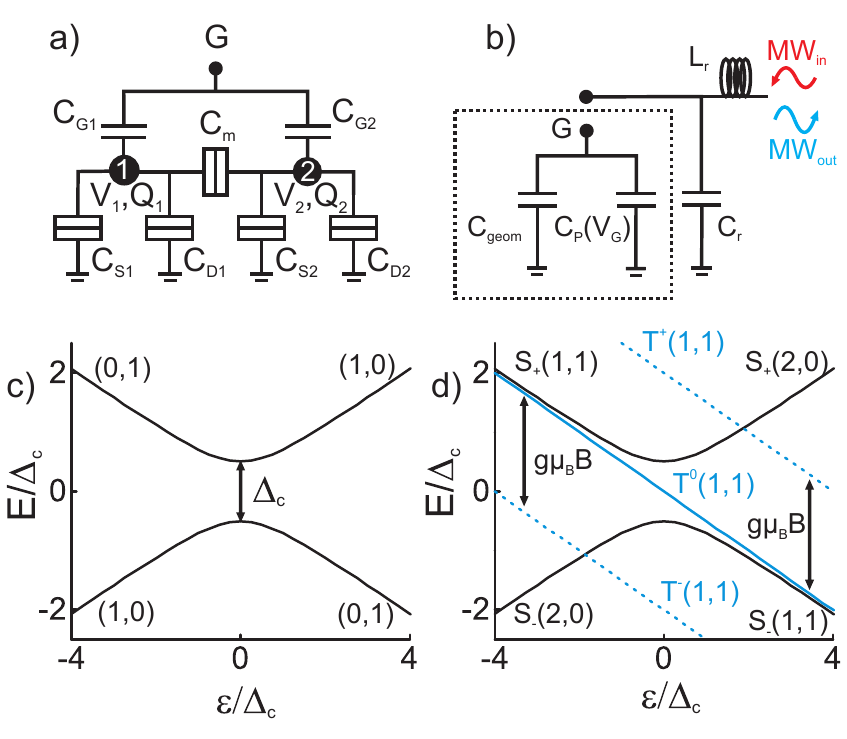}
\caption{Charge and spin qubits in double quantum dots. (a) A schematic of a DQD system composed of QDs 1 and 2 coupled to a gate electrode G by capacitances $C_\text{G1}$ and $C_\text{G2}$, respectively. (b) Equivalent circuit of (a) including the geometrical and parametric capacitance in parallel. The resonator, with characteristic inductance $L_\text{r}$ and capacitance to ground $C_\text{r}$ creates an oscillatory voltage on the gate of the DQD. (c) Energy level spectrum of a DQD as a function of detuning in the single electron regime. Mixing of the states forms a finite gap of magnitude $\Delta_\text{c}$ at $\varepsilon=0$. (d) The energy level spectrum of a DQD in the two-electron regime. Split-off triplet states, T$^{0}$, T$^{-}$ and T$^{+}$ must be considered in addition to the mixed singlet states.}
\label{fig:1}
\end{figure}

We consider a DQD system as illustrated in Fig.~\ref{fig:1}(a), in which the two quantum dots (QDs) 1 and 2 are coupled to a sensing gate electrode (G) via gate capacitances \(C_{\text{G}i}\) for $i=1,2$. Each QD is further coupled to source and drain electrodes by capacitances \(C_{\text{S}i}\) and \(C_{\text{D}i}\) respectively and they are further coupled to each other by a mutual capacitance $C_\text{m}$. Here, we study the zero-bias regime in which source and drain electrodes are connected to ground.

The DQD has an associated differential capacitance, \(C_\text{diff}\), as seen from the gate that can be obtained from the following expression

\begin{equation}
\label{eq:4}
C_\text{diff}=\frac{d(Q_1+Q_2)}{dV_\text{G}}.
\end{equation}

We can express the net charges $Q_{i}$ in QD $i$ by considering the total existing charge in the islands $-e\left\langle n_i\right\rangle$, and the polarization charge induced by the gate electrode and the mutual electrostatic effect between dots

\begin{gather}
\begin{aligned}
Q_{1} &= C_\text{G1}(V_{G}-V_{1}) \\
Q_{1} &= -e\left\langle n_{1}\right\rangle+(C_\text{S1}+C_\text{D1})V_{1}-C_\text{m}(V_{2}-V_{1}) \\
Q_{2} &= C_\text{G2}(V_{G}-V_{2}) \\
Q_{2} &= -e\left\langle n_{2}\right\rangle+(C_\text{S2}+C_\text{D2})V_{2}-C_\text{m}(V_{1}-V_{2}).
\end{aligned}
\label{eq:5}
\end{gather}

Here $V_i$ and $\left\langle n_i\right\rangle$ are the voltage and average electron number in QD $i$ and $e$ is the electron charge. Solving Eqs.~\eqref{eq:5} for $Q_{1}+Q_{2}$ through elimination of $V_{1}$ and $V_{2}$, we arrive at the simplified expression 

\begin{equation}
\begin{split}
\begin{aligned}
Q_1+Q_2 &=  e (\alpha_{1}\left\langle n_{1}\right\rangle+\alpha_{2}\left\langle n_{2}\right\rangle) \\
&+ [\alpha_{1}(C_\text{S1}+C_\text{D1})+\alpha_{2}(C_\text{S2}+C_\text{D2})]V_{G} 
\end{aligned}
\end{split}
\label{eq:6}
\end{equation}

in the limit of $C_{m} \ll C_{\Sigma i}$. Here $C_{\Sigma i}=C_{\text{S}i}+C_{\text{D}i}+C_{\text{G}i}+C_\text{m}$ is the total capacitance of QD $i$ and $\alpha_i$ its gate coupling defined as \(C_{\text{G}i}\)/\(C_{\Sigma i}\). Inserting Eq.~\eqref{eq:6} into Eq.~\eqref{eq:4} gives the differential capacitance as the sum of two components

\begin{equation}
C_\text{diff}=C_\text{geom} + C_\text{p}.
\label{eq:7}
\end{equation}

The \textit{geometrical capacitance}, $C_\text{geom}$, a voltage-independent term, represents the DC limit of the capacitance

\begin{equation}
C_{geom}=\alpha_{1}(C_\text{S1}+C_\text{D1})+\alpha_{2}(C_\text{S2}+C_\text{D2}).
\label{eq:8}
\end{equation}

Here, the total geometrical capacitance appears as the sum of the gate capacitance in series with the parallel capacitances of the source and drain tunnel junctions for each QD.  

The second term in Eq.~\eqref{eq:7}, $C_\text{p}$, represents the parametric dependence of the time-averaged excess electron numbers on the gate voltage, and is thus coined the \textit{parametric capacitance}. This is written as

\begin{equation}
C_\text{p}=-e\left[\alpha_2\dfrac{\partial \langle n_{2} \rangle}{\partial V_{G}}+\alpha_1\dfrac{\partial \langle n_{1} \rangle}{\partial V_{G}}\right]=-e\alpha^{'} \dfrac{\partial \langle n_{2} \rangle}{\partial V_\text{G}}
\label{eq:10}
\end{equation}

since $\left\langle n_{1}\right\rangle=-\left\langle n_{2}\right\rangle$ when QDs exchange electrons and \(\alpha^{'}=(\alpha_{2}-\alpha_{1})\). The equivalent circuit of the DQD at finite frequencies is depicted inside the dashed rectangle of Fig.~\ref{fig:1}(b) where the geometrical and parametric capacitance appear in parallel between electrode G and ground. The simplicity of this equivalent circuit can be exploited to obtain information of the charge and spin state of the DQD. We focus on the situation in which the state of the DQD is probed dispersively at finite-frequencies by monitoring the state of a resonator coupled to the system via the gate electrode G, as shown in Fig.~\ref{fig:1}(b). Here, we depict a lumped-element resonator composed by an inductance \(L_\text{r}\) and a capacitance to ground \(C_\text{r}\) and hence with a natural frequency of resonance given by $f_\text{r}=\omega_\text{r}/2\pi=1/2\pi\sqrt{L_\text{r}C_\text{r}}$. However, for the discussion hereinafter, the argument applies for any resonator, such as distributed electrical resonators~\cite{Petersson2012,Shevchenko2012} and mechanical resonators~\cite{LeHaye2009,Shevchenko2012b}. When the resonator is probed at a finite frequency $f_\text{r}$, it produces an oscillatory gate voltage of amplitude $\delta V_\text{G}$ given by 

\begin{equation}
V_{G}(t)=\delta V_{G}\text{sin}(2\pi f_rt),
\label{eq:1}
\end{equation}

which in turn produces a redistribution of charges $Q_1$ and $Q_2$ in the DQD due to the associated differential capacitance in Eq.~\eqref{eq:7}. This additional state-dependent capacitance loads the resonator leading to a dispersive shift of the resonant frequency. This frequency shift can be readily monitored using standard homodyne detection techniques, such as reflectometry~\cite{Schoelkopf1998, Colless2013, Gonzalez-Zalba2014, Gonzalez-Zalba2015, Hile2015} or transmission~\cite{Wallraff2004, Frey2012}. In the following sections, we present a description of the DQD-resonator interaction. We explore the DQD in the charge and spin qubit configurations and exploit the dependence of the parametric capacitance on external variables for qubit characterization.

\section{III. Charge Qubit}

Operating in the single electron regime, where one electron is shared between the QDs, the DQD will function as a charge qubit~\cite{Kim2015}. Two discrete charge states are considered, corresponding to the configurations $(n_1,n_2)=(1,0)$ and (\(0,1\)). The Hamiltonian of such a two-level system in terms of Pauli spin matrices $\sigma_{x,y,z}$ is 

\begin{equation}
H=-\dfrac{\Delta_\text{c}}{2}\sigma_{x} - \dfrac{\varepsilon}{2}\sigma_{z},
\label{eq:2}
\end{equation}

where $\varepsilon$ represents the energy detuning between quantum dots and $\Delta_\text{c}$ the tunnel coupling that mixes the (1,0) and (0,1) states, creating a point of avoided crossing at $\varepsilon=0$. The corresponding eigenenergies, as illustrated in Fig.~\ref{fig:1}(c), are given by

\begin{equation}
E_{\pm}=\pm \dfrac{1}{2} \sqrt{\varepsilon^{2}+\Delta_\text{c}^{2}}.
\label{eq:3}
\end{equation}

We can calculate the parametric capacitance of a charge qubit by taking into account that an oscillatory voltage on the gate will induce an oscillatory variation on the DQD energy detuning centered around $\varepsilon_0$ as $\varepsilon(t)=\varepsilon_0+\delta\varepsilon\sin(2\pi f_\text{r}t)$, where $\delta\varepsilon=-e\alpha^{'}\delta V_\text{G}$. Then Eq.~\eqref{eq:10} becomes

\begin{equation}
C_\text{p}=(e\alpha^{'})^2\dfrac{\partial \langle n_{2} \rangle}{\partial\varepsilon}
\label{eq:11}
\end{equation}

and the problem simplifies to calculating $\left\langle n_2\right\rangle(\varepsilon)$.

\subsection{A. Theory}

In the energy basis, the time-averaged excess electron number in QD2 in the single electron regime may be expressed as

\begin{equation}
\langle n_{2} \rangle = \left\langle n_2\right\rangle_{-}P_{-} + \left\langle n_2\right\rangle_{+}P_{+},
\label{eq:13}
\end{equation}

where $P_{\pm}$ are the probabilities of the excess electron occupying the ground(-) or excited(+) state and 

\begin{equation}
\left\langle n_2\right\rangle_{\pm} = \frac{1}{2} \left(1 \pm \dfrac{\varepsilon}{\Delta E} \right)
\label{eq:14}
\end{equation}

are the average number of electrons in QD2 in the ground and excite state~\cite{Gonzalez-Zalba2016}. Here, $\Delta E=E_+-E_-=\sqrt{\varepsilon^2 + \Delta_\text{c}^{2}}$. Using Eqs.~\eqref{eq:11}-\eqref{eq:14}, we can then obtain the instantaneous parametric capacitance of a charge qubit:


\begin{equation}
C_\text{p}(t)=C_{0} \Biggl( 
\underbrace{\dfrac{\Delta_\text{c}^{3}}{\Delta E^{3}}\chi_\text{c} \rule[-15pt]{0pt}{1.7pt}}_{\mbox{\footnotesize quantum}} +
\underbrace{\dfrac{\varepsilon \Delta_\text{c}}{\Delta E} \dfrac{\partial\chi_\text{c}}{\partial t}\frac{\partial t}{\partial\varepsilon} \rule[-15pt]{0pt}{1.7pt}}_{\mbox{\footnotesize tunnelling}},
\Biggr)
\label{eq:15}
\end{equation}

where $C_{0}= (e\alpha^{'})^2/2\Delta_\text{c}$, $\chi_\text{c}=P_{-}-P_{+}$, the difference between ground and excited state occupation probabilities and we have expanded $\frac{\partial\chi_\text{c}}{\partial\varepsilon}=\frac{\partial\chi_\text{c}}{\partial t}\frac{\partial t}{\partial\varepsilon}$. Eq.~\eqref{eq:15} contains two detuning-dependent contributions. The first term, corresponds to the so-called {\itshape quantum capacitance} arising from adiabatic transitions in systems with finite curvature of the energy bands~\cite{Sillanpaa2005,Duty2005}. The second term, the {\itshape tunnelling capacitance}, appears when non-adiabatic processes, such as relaxation and thermal or resonant excitation, occur at a rate comparable or faster than the probing frequency $f_\text{r}$~\cite{Ashoori1992,Ciccarelli2011,Gonzalez-Zalba2015, Gonzalez-Zalba2016}. 

Typically, in reflectometry measurements, $C_\text{p}(t)$ is averaged over a resonator cycle. Hence, calculating the averaged parametric capacitance $\left\langle C_\text{p}\right\rangle$ requires a time-dependent solution of the probability difference $\chi_\text{c}$. This can be done by analysing the dynamics of the driven two-level system which is given by the following master equation

\begin{equation}\label{master}
 \begin{aligned}
	\dot{P_-}&=&\Gamma_-P_+-\Gamma_+P_- \\
	\dot{P_+}&=&-\Gamma_-P_++\Gamma_+P_-,
 \end{aligned}
\end{equation}

where $\Gamma_-=1/T_1^\text{c}$ is the relaxation rate from the excited state to the ground state and $\Gamma_+=\Gamma_-e^{-\Delta E/k_\text{B}T}$ is the relaxation rate from the ground state to the excited state~\cite{Berns2006}. The solution to first order approximation with respect to small variations of detuning $\delta\varepsilon\ll\Delta_c$, can be calculated following references~\cite{Persson2010a,Gonzalez-Zalba2015}. In this approximation, the probability difference can be expressed as $\chi_\text{c}(t)=\chi_\text{c}^0+\delta\chi_\text{c}(t)$ where $\delta\chi_\text{c}(t)$ represents a small deviation from the equilibrium occupation probability difference

\begin{equation}
\begin{aligned}
\chi_\text{c}^0 &= \dfrac{1}{Z_\text{c}^0}  \Big( e^{\Delta E(\varepsilon_0)/2k_\text{B}T} - e^{-\Delta E(\varepsilon_0)/2k_\text{B}T} \Big) \\
&= \mathrm{tanh}(\Delta E(\varepsilon_0)/2k_\text{B}T),
\end{aligned}
\label{eq:16}
\end{equation}

where $Z_\text{c}^0 = \mathrm{exp}[\Delta E(\varepsilon_0)/2k_\text{B}T] +  \mathrm{exp}[-\Delta E(\varepsilon_0)/2k_\text{B}T]$ is the partition function for the charge qubit. We solve Eq.~\ref{master} for $\delta\chi_\text{c}(t)$ and obtain the following expression

\begin{equation}\label{dchi}
	\delta\chi_\text{c}(t)=-\frac{2\eta}{\omega_\text{r}^2+\gamma^2}\left[\gamma\sin(\omega_\text{r}t)-\omega_{r}\cos(\omega_\text{r}t)\right]\delta\varepsilon
\end{equation}

where $\eta$ and $\gamma$ are temperature-dependent coefficients given by

\begin{equation}\label{eta}
 \begin{aligned}
		\eta&=-\frac{1}{T_1^\text{c}}\frac{\varepsilon_0}{k_\text{B}T\Delta E(\varepsilon_0)}\frac{e^{-\Delta E(\varepsilon_0)/2k_\text{B}T}}{e^{\Delta E(\varepsilon_0)/2k_\text{B}T}+e^{-\Delta E(\varepsilon_0)/2k_\text{B}T}} \\
		\gamma&=\frac{1}{T_1^\text{c}}\left(1+e^{-\Delta E(\varepsilon_0)/k_\text{B}T}\right).
 \end{aligned}
\end{equation}

Calculating the instantaneous parametric capacitance to first order approximation in $\delta\varepsilon$ using Eqs.~(\ref{eq:16})-(\ref{eta}) and averaging over a resonator cycle we obtain an analytical expression for parametric capacitance of a charge qubit

\begin{equation}\label{CQpara}
	\begin{aligned}
	&\left\langle C_\text{p}\right\rangle\approx C_0\left\{\frac{\Delta_\text{c}^3}{\Delta E^3(\varepsilon_0)}\chi_\text{c}^0\right. + \\
	&+\frac{2\Delta_\text{c}}{k_\text{B}T}\frac{\varepsilon_0^2}{\Delta E(\varepsilon_0)^2}\left[g(\varepsilon_0,T)-1)\right]\frac{\Gamma_1^\text{c2}}{\omega_\text{r}^2+g^2(\varepsilon_0,T)\Gamma_1^\text{c2}}\biggl\}
	\end{aligned}
\end{equation}


where $\Gamma_1^\text{c}=1/T_1^\text{c}$ and $g(\varepsilon_0,T)=1+e^{-\Delta E(\varepsilon_0)/k_\text{B}T}$.

\subsection{B. Results}

\subsubsection{Slow and Fast Relaxation Limits}

\begin{figure}[t]
\centering
\includegraphics{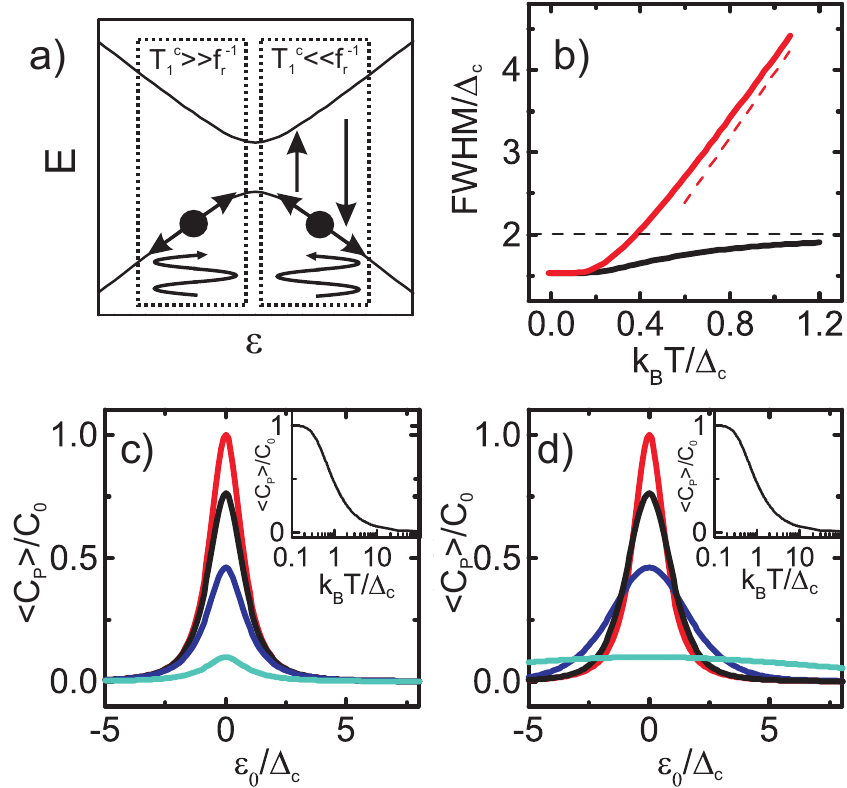}
\caption{Charge Qubit. (a) Schematic energy level diagram of the DQD indicating the electron dynamics in the slow ($\Gamma_{1}^\text{c}\ll \omega_\text{r}$) and fast ($\Gamma_{1}^\text{c}\gg \omega_\text{r}$) relaxation regime. Only in the fast relaxation regime the system can track the instantaneous thermal population of the system exemplified by the vertical black arrows. (b) FWHM of the parametric capacitance lineshape in the slow (black solid line) and fast relaxation regime (red solid line). The slow regime tends asymptotically to FWHM$=2\Delta_\text{c}$ (black dashed line) and the fast regime to FWHM$=3.53k_\text{B}T$ (red dahsed line). Parametric capacitance lineshape as a function of detuning for $k_\text{B}T/\Delta_\text{c}= 0.1, 0.5, 1, 2$ (red, black, dark blue, light blue solid lines) in the slow (c) and fast (d) relaxation regimes. Insets indicate the maximum parametric capacitance as a function of increasing temperature for both regimes.}
\label{fig:2}
\end{figure}

In the limit, $\Gamma_{1}^\text{c} \ll \omega_\text{r}$, in which the charge relaxation time is longer than the period of the resonator, thermal repopulation does not occur on the time-scale of the resonator and hence $\partial\chi_\text{c}/\partial t=0$, represented schematically on the left side of Fig.~\ref{fig:2}(a). In such cases, the parametric capacitance will be composed solely of quantum capacitance contributions. 

\begin{equation}
	\begin{aligned}
	\left\langle C_\text{p}\right\rangle\approx & C_0\frac{\Delta_\text{c}^3}{\Delta E^3(\varepsilon_0)}\chi_\text{c}^0
	\end{aligned}
\end{equation}

Conversely, for $\Gamma_{1}^\text{c} \gg \omega_\text{r}$, the thermal population distribution tracks the instantaneous excitation from the resonator ($\partial\chi_\text{c}/\partial\varepsilon=\partial\chi_\text{c}^0/\partial\varepsilon_0$) as schematically depicted in the right side of Fig.~\ref{fig:2}(a) and therefore both quantum and tunnelling capacitances must be considered

\begin{equation}
	\begin{aligned}
	\left\langle C_\text{p}\right\rangle\approx C_0\left\{\frac{\Delta_\text{c}^3}{\Delta E^3(\varepsilon_0)}\chi_\text{c}^0 + \dfrac{\varepsilon_0\Delta_\text{c}}{\Delta E(\varepsilon_0)} \dfrac{\partial\chi_\text{c}^0}{\partial\varepsilon_0}\right\}.
	\end{aligned}
\end{equation}

In Fig.~\ref{fig:2}(b-d), we present a lineshape analysis of the parametric capacitance of a charge qubit as a function of temperature. The slow (Fig.~\ref{fig:2}(c)) and fast-relaxation regimes (Fig.~\ref{fig:2}(d)) exhibit a symmetric capacitance lineshape with a maximum at the avoided crossing point (\(\varepsilon_{0}=0\)). Here, we plot the lineshape for increasing temperature from $k_\text{B}T/\Delta_\text{c}=0.1$ (red line) to $k_\text{B}T/\Delta_\text{c}=5$ (light blue line). The maximum parametric capacitance diminishes with increasing temperature according to \(\mathrm{tanh}(\Delta_\text{c}/2k_\text{B}T)\) as can be seen in both insets and it becomes vanishingly small for values $k_\text{B}T/\Delta_\text{c}>10$. This is found to be identical under both regimes due to the tunnelling capacitance having no contribution at the avoided crossing. Moreover, in the low temperature limit both regimes show a full width at half maximum, FWHM $=1.53\Delta_\text{c}$, see in Fig.~\ref{fig:2}(b). However, a distinction between the two regimes may be made through contrasting the temperature dependence of the FWHM for $k_\text{B}T/\Delta_\text{c}>0.2$. In the slow-relaxation limit the FWHM increases until it saturates to a value $2\Delta_\text{c}$ at high temperatures. Conversely, in the fast-relaxation regime, the FHWM shows an increase with temperature due to the tunnelling capacitance. This increase tends asymptotically to a linear dependence with $T$ at high temperatures, FWHM$=3.53k_\text{B}T$ (red dashed line). This different behaviour of the FWHM as a function of temperature allows identifying the relaxation regime in which the charge qubit is. Moreover, at low temperatures, a fit to the FWHM allows measuring $\Delta_{c}$ independently of the relaxation regime in which the qubit is. Finally, the fast relaxation regime at high temperatures allows calibrating the electron temperature of the system.

\subsubsection{Intermediate Relaxation Regime}

\begin{figure}[htbp]
	\centering
		\includegraphics{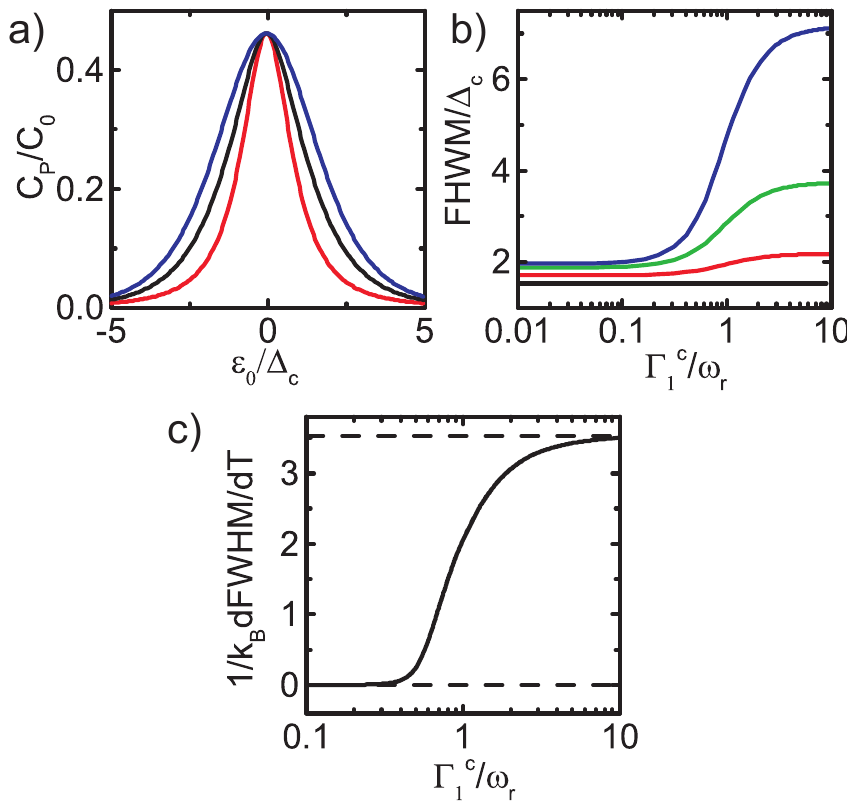}
	\caption{Charge qubit relaxation. (a) Parametric capacitance lineshape as a function of detuning for $\Gamma_1^\text{c}/\omega_\text{r}=0, 1$ and 10 (red, black, blue lines).}
	\label{Fig2b}
\end{figure}

In the limit, $\Gamma_{1}^\text{c} \approx \omega_\text{r}$, in which the charge relaxation time is similar to the period of the resonator, the tunnelling capacitance term contributes partially and the full Eq.~\ref{CQpara} must be used. This regime has been observed in superconducting qubits~\cite{Persson2010a} and silicon quantum dots~\cite{Gonzalez-Zalba2015}. As can be seen in Fig.~\ref{Fig2b}(a), the effect of increasing $\Gamma_{1}^\text{c}$ is to increase the FWHM of the capacitance peak going from the slow relaxation limit $\Gamma_1^\text{c}/\omega_\text{r}=0$ (red trace) to the fast relaxation regime $\Gamma_1^\text{c}/\omega_\text{r}=10$ (blue trace). The FWHM increase as a function of $\Gamma_1^\text{c}/\omega_\text{r}$, Fig.~\ref{Fig2b}(b), is proportional to the temperature of the system, going from no change at low temperatures (black line at$k_\text{B}T/\Delta_\text{c}=0.1$) to FWHM $\approx 3.53k_\text{B}T$ at high temperatures (blue line at $k_\text{B}T/\Delta_\text{c}=2$). Finally, in Fig.~\ref{Fig2b}(c), we explore the possibility of determining $\Gamma_{1}^\text{c}$ from a measurement of the FWHM vs $k_\text{B}T$ slope at high temperatures. As can be seen in Fig.~\ref{Fig2b}(c), the slope $d$FWHM/$dT$, varies from 0 at the slow relaxation limit to $3.53k_\text{B}$ in the fast relaxation regime. Hence a fit to the linear region of a FWHM vs $T$ plot allows a direct measurement of $\Gamma_1^\text{c}/\omega_\text{r}$.

\section{IV. Spin Qubits}

We now move on to the discussion of the DQD in the two-electron regime. The electron configurations read \((1,1)\) and \((2,0)\). In addition to the singlet states ground(excited) S$_{-(+)}(1,1)$ and S$_{-(+)}(2,0)$, the \((1,1)\) configuration introduces a set of degenerate triplet states, \(\mathrm{T}^{-}, \mathrm{T}^{0}\) and \(\mathrm{T}^{+}\), whose degeneracy is broken upon application of a magnetic field. The energies of the triplet states, as shown in Fig.~\ref{fig:1}(d), are given by

\begin{gather}
\begin{aligned}
E_{T^0} &= -\varepsilon/2\\
E_{T^\pm} &=  -\varepsilon/2\pm g\mu_BB, \\
\end{aligned}
\label{eq:triplets}
\end{gather}

where $\mu_B$ is the Bohr magneton and $g$ the electron g-factor. The transitions between the two states are dictated by the Pauli spin selection rules~\cite{Betz2015} and hence only charge states with same spin numbers couple. We neglect here spin-orbit fields and the small singlet-triplet coupling typically due to time-varying hyperfine fields in semiconductor QDs~\cite{Petta2010, Maune2012}. Any mixing of the T\((1,1)\) and T\((2,0)\) states is assumed to occur at large detunings away from the singlet avoided crossing at \(\varepsilon=0\) due to the typically large orbital energy splitting in QDs~\cite{Petta2010, Lim2011, Betz2015}. Therefore, their energies present a linear dependence with respect to detuning. The DQD in the two electron regime may function as a spin qubit~\cite{Veldhorst2015} or a singlet-triplet qubit~\cite{Petta2005, Maune2012, Wu2014}.

\subsection{A. Theory}

Beginning in the energy basis, the average excess electron number in the two-electron regime is written as 
\begin{equation}
\langle n_{2} \rangle = \left\langle n_2\right\rangle_{-}P_{-}^\text{s} + \left\langle n_2\right\rangle_{+}P_{+}^\text{s} + \left\langle n_2\right\rangle^\text{t}P^\text{t},
\label{eq:17}
\end{equation}

where $\left\langle n_2\right\rangle^\text{t}$ and $P^\text{t}$ are the averaged excess electron number in the triplet states and the corresponding occupation probability and $P_{-(+)}^\text{s}$ are the singlet ground(-) and excited(+) state probabilities respectively. Due to the Pauli exclusion principle, $\left\langle n_2\right\rangle^\text{t}=1$ and the normalized probabilities require $P_{-}^\text{s}+P_{+}^\text{s}+P^\text{t}=1$. The average number of electron in QD2 in the spin configuration can be written as

\begin{equation}
\langle n_{2} \rangle = 1-\dfrac{1}{2}\chi_\text{s}^{'}+\dfrac{\varepsilon}{2\Delta E}\chi_\text{s}.
\label{eq:18}
\end{equation}

Here $\chi_\text{s}=P_{-}^\text{s}-P_{+}^\text{s}$ and $\chi_\text{s}^{'}=P_{-}^\text{s}+P_{+}^\text{s}$. From Eq.~\eqref{eq:11} this gives the instantaneous parametric capacitance of the spin qubit system

\begin{equation}
C_\text{p}(t)=C_{0} \Biggl( 
\underbrace{ \dfrac{\Delta_\text{c}^{3}}{\Delta E^{3}}\chi_\text{s} \rule[-15pt]{0pt}{5pt}}_{\mbox{\footnotesize quantum}} + 
\underbrace{ \dfrac{\varepsilon\Delta_\text{c}}{\Delta E} \dfrac{\partial\chi_\text{s}}{\partial \varepsilon} - \Delta_\text{c} \dfrac{\partial\chi_\text{s}^{'}}{\partial \varepsilon} \rule[-15pt]{0pt}{5pt}}_{\mbox{\footnotesize tunnelling}}
\Biggr).
\label{eq:19}
\end{equation}

We note this expression is similar to Eq.~\ref{eq:15} for the charge qubit with an additional term $-\Delta_\text{c}\partial\chi_\text{s}^{'}/\partial \varepsilon$ in the tunnelling capacitance and the different occupation probabilities now distributed over five spin states. 

%
%
%

In the following, we introduce the relaxation rate between states with different spin state, $\Gamma_1^\text{s}=(T_1^\text{s})^{-1}$, where $T_1^\text{s}$ is the spin relaxation time.

\subsection{B. Results}

\subsubsection{Slow Relaxation Regime}

Following a similar analysis to the charge qubit in the slow charge relaxation limit, when considering $\Gamma_{1}^\text{c}, \Gamma_{1}^\text{s} \ll \omega_{r}$, the tunnelling capacitance can be neglected and the parametric capacitance is solely composed of quantum capacitance contributions

\begin{equation}
	\begin{aligned}
	\left\langle C_\text{p}\right\rangle\approx & C_0\frac{\Delta_\text{c}^3}{\Delta E^3(\varepsilon_0)}\chi_\text{s}^0,
	\end{aligned}
\end{equation}

where $\chi_s^0$ is the equilibrium probability distribution for the spin qubit

\begin{equation}
	\chi_\text{s}^0 = \dfrac{1}{Z_\text{s}^0} \Big(e^{\Delta E(\varepsilon_0)/2k_\text{B}T} -  e^{-\Delta E(\varepsilon_0)/2k_\text{B}T} \Big)
\end{equation}

and for which the new partition function, $Z_\text{s}^0$, incorporates the additional three triplet states

\begin{equation}
\begin{aligned}
Z_\text{s}^0 = \quad & e^{\Delta E(\varepsilon_0)/2k_\text{B}T} +  e^{-\Delta E(\varepsilon_0)/2k_\text{B}T}+ \\
& e^{\varepsilon_0/2k_\text{B}T} \lbrace 1+ e^{-g\mu_{B}B/k_\text{B}T}+e^{g\mu_{B}B/k_\text{B}T} \rbrace.
\end{aligned}
\label{eq:21}
\end{equation}

\begin{figure}
\centering
\includegraphics{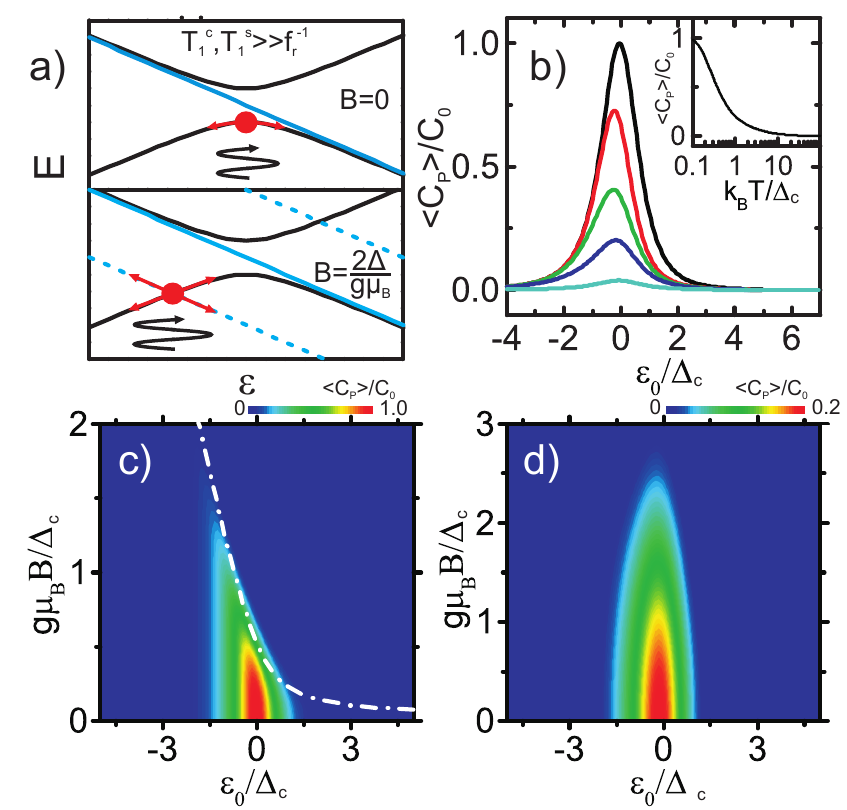}
\caption{Spin qubit in the slow relaxation regime.  (a) Schematic of the electron dynamics at $B=0$ (top panel) and $B=2\Delta_\text{c}/g\mu_B$ (bottom panel). At zero field, the electron tracks the curvature of the singlet ground state. At finite fields the electron performs diabatic transitions at the singlet-triplet crossing and no relaxation occurs in the timescale of the resonator. (b) Parametric capacitance lineshape as a function of detuning for $k_\text{B}T/\Delta_\text{c}= 0.1, 0.25, 0.5, 1.0, 5$ (black, red, green, dark blue, light blue solid lines). The insets indicates the maximum parametric capacitance as a function of increasing temperature. Parametric capacitance as a function of detuning and magnetic field at $k_\text{B}T/\Delta_\text{c}=0.1$ (c) and $k_\text{B}T/\Delta_\text{c}=1$ (d). The dot-dashed line in (c) indicates the position in $\varepsilon-B$ plane of the singlet-triplet crossing point.}
\label{fig:3}
\end{figure}

This is the case, for example, for singlet-triplet systems in silicon DQDs and coupled dopant-dot~\cite{Betz2015,Urdampilleta2015,House2015}. In Fig.~\ref{fig:3}, we show the DQD energy levels for two different magnetic fields. In the upper panel, $B=0$, the triplet states are degenerate and the ground state of the system around $\varepsilon=0$ is a singlet. For large positive detunings the S$_-(1,1)$ and T$(1,1)$ are degenerate. Here a quantum capacitance signal arises from the finite curvature of the singlet branch. The lineshape is symmetrical for $k_\text{B}T/\Delta_\text{c}\leq0.1$ with a FWHM $= 1.53\Delta_\text{c}$ but develops a slight asymmetry as the temperature is increased from $k_\text{B}T/\Delta_\text{c}=0.1$ (black line) to $k_\text{B}T/\Delta_\text{c}=5$ (light blue line) as seen in Fig.~\ref{fig:3}(b). A reduction of signal towards positive detunings is observed due to the repopulation of the triplet branch that possess no quantum capacitance contribution due to its lack of band curvature. Moreover, the maximum capacitance decreases with increasing temperature, as seen in the inset.

We now explore the magnetic field-dependence of the parametric capacitance of the system. At finite fields, depicted in the bottom panel of Fig.~\ref{fig:3}(a), the triplet degeneracy is broken by the Zeeman energy and the T$^{-}$ state becomes the new ground state of the system for $\varepsilon>\left[\Delta_\text{c}^2-(2g\mu_BB)^2\right]/4g\mu_BB$. Since no tunnelling is allowed between the S$_{-}$ and T$^{-}(1,1)$ in the time-scale of the resonator, as the magnetic field is increased, the capacitance vanishes. This can be seen in Fig.~\ref{fig:3}(c,d) for $k_\text{B}T/\Delta_\text{c}=0.1$ and 1 respectively, where we plot $\left\langle C_\text{p}\right\rangle/C_0$ as a function of magnetic field and detuning. For $k_\text{B}T/\Delta_\text{c}=0.1$ the loss of signal occurs asymmetrically tracking the singlet-triplet crossing point (white dashed line) and becoming vanishingly small for $g\mu_BB/\Delta_\text{c}>1.5$. Conversely, for $k_\text{B}T/\Delta_\text{c}=1$, the signal vanishes quasi-symmetrically and a larger magnetic field is necessary to observe the loss of capacitance. Overall, measuring at low temperature the magnetic-field dependence of the parametric capacitance in slow-relaxation systems allows determining relevant qubit parameters: $\Delta_\text{c}$ from the FWHM in the low temperature limit, $T$ from the decay of the capacitance signal as a function of magnetic field at fixed detuning and $g$ from the shift in detuning of the capacitance maximum as a function of $B$.

\subsubsection{Fast Relaxation Regime}

\begin{figure}[t]
\centering
\includegraphics{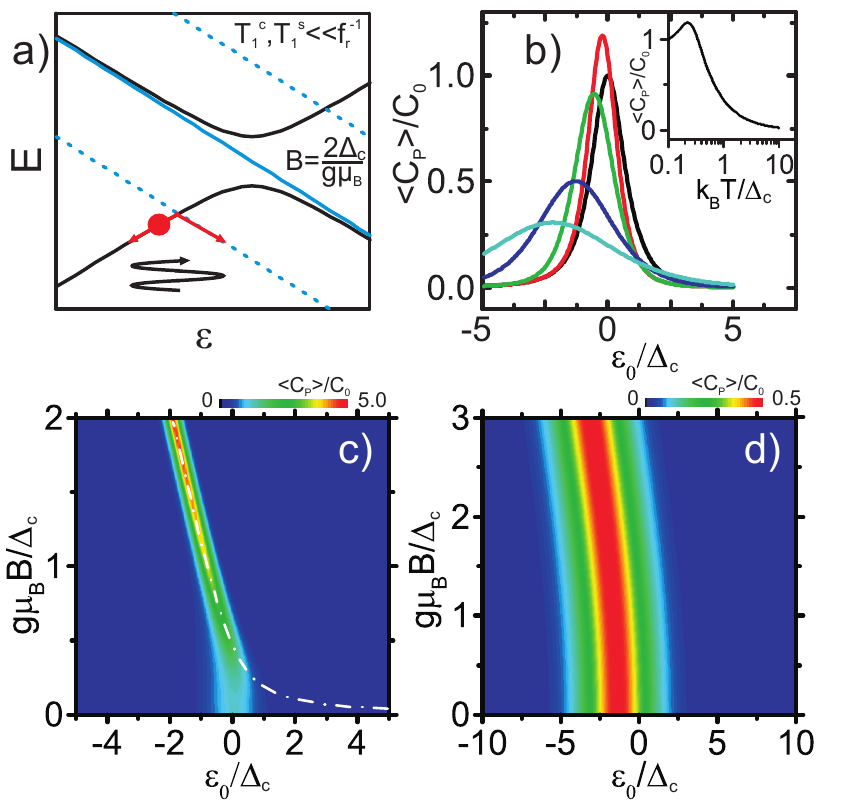}
\caption{ Spin qubit in the fast relaxation regime. (a) Schematic of the electron dynamics at $B=2\Delta_\text{c}/g\mu_B$. The electron performs adiabatic transitions at the singlet-triplet crossing since relaxation occurs much faster than the timescale of the resonator drive.  (b) Parametric capacitance lineshape as a function of detuning for $k_\text{B}T/\Delta_\text{c}= 0.1, 0.3, 0.5, 1.0, 1.65$ (black, red, green, dark blue, light blue solid lines). The insets indicates the maximum parametric capacitance as a function of increasing temperature. Parametric capacitance as a function of detuning and magnetic field at $k_\text{B}T/\Delta_\text{c}=0.1$ (c) and $k_\text{B}T/\Delta_\text{c}=1$ (d). The white dot-dashed line in (c) indicates the singlet-triplet crossing point.}
\label{fig:4}
\end{figure}

In this section, we explore the singlet-triplet system in the fast relaxation limit $\Gamma_{1}^\text{c}, \Gamma_{1}^\text{s}\gg \omega_\text{r}$ where the system tracks the instantaneous thermal distribution imposed by oscillatory value of the detuning. In this scenario, tunnelling between the S$_{-}(2,0)$ and T$^{-}(1,1)$ occurs in the time-scale of the resonator, as depicted in Fig.~\ref{fig:4}(a), giving a finite tunnelling capacitance contribution. In this limit the parametric capacitance reads,

\begin{equation}
	\begin{aligned}
	\left\langle C_\text{p}\right\rangle\approx C_0 \left\{\frac{\Delta_\text{c}^3}{\Delta E^3(\varepsilon_0)}\chi_\text{s}^0+\frac{\varepsilon_0\Delta_\text{c}}{\Delta E(\varepsilon_0)}\frac{\partial\chi_\text{s}^0}{\partial \varepsilon_0} - \Delta_\text{c}\dfrac{\partial\chi_\text{s}^{'0}}{\partial\varepsilon_0}\right\}
	\end{aligned}
\end{equation}

where $\chi_\text{s}^{'0}$ corresponds to

\begin{equation}
	\chi_\text{s}^{'0} = \dfrac{1}{Z_\text{s}^0} \Big(e^{\Delta E(\varepsilon_0)/2k_\text{B}T} +  e^{-\Delta E(\varepsilon_0)/2k_\text{B}T} \Big)
\end{equation}

In Fig.~\ref{fig:4}(b), we demonstrate the additional effect of the tunnelling capacitance on the temperature dependence of the lineshape from $k_\text{B}T/\Delta_\text{c}=0.1$ (black line) to $k_\text{B}T/\Delta_\text{c}=1.65$ (light blue line).

We observe several differences with the slow relaxation regime that allows identifying the fast relaxation limit. First of all, we see that the lineshape is symmetrical around $\varepsilon_0=0$ and shifts towards negative values of detuning as the temperature increases with a linear dependence given by $\left|\varepsilon_0\right|\cong 1.43k_\text{B}T$. Additionally, we observe an increase of the maximum parametric capacitance as the temperature is increased to $k_\text{B}T/\Delta_\text{c}\cong 0.225$ and then a rapid decrease until it becomes vanishingly small for $k_\text{B}T/\Delta_\text{c}>10$ (see Inset). Finally, we calculate a FWHM$=1.53\Delta_\text{c}$ in the low temperature limit that tends asymptotically to FWHM$=3.53k_\text{B}T$ at high temperatures.

The magnetic field dependence of the parametric capacitance also reveals additional differences when compared to the slow relaxation regime. Here the capacitance signal does not vanish with increasing $B$, see Fig.~\ref{fig:4}(c,d). At low temperatures ($k_\text{B}T/\Delta_\text{c}=0.1$), we observe a transition between a quantum capacitance dominated signal for $g\mu_BB/\Delta_\text{c}<0.5$, since S$_-$ is the ground state at $\varepsilon=0$, to a tunnelling capacitance signal for $g\mu_BB/\Delta_\text{c}>0.5$. In this situation, the maximum $\left\langle C_\text{p}\right\rangle/C_{0}$ tends to $\Delta_\text{c}/2k_\text{B}T$ for large magnetic fields and tracks the position of the singlet-triplet crossing point (white dot-dashed line) as seen in Fig.~\ref{fig:4}(c). At higher temperatures, $k_\text{B}T/\Delta_\text{c}=1$ in Fig~\ref{fig:4}(d), the maximum $\left\langle C_\text{p}\right\rangle/C_{0}$ also tracks the position of the singlet-triplet crossing but in this case, the signal is dominated by the tunnelling capacitance contribution for all values of $B$. Overall, we can exploit the magnetic-field dependence of the parametric capacitance in the low temperature limit to determine $\Delta_\text{c}$ from the FWHM in the low temperature limit or from the magnetic field value at which the maximum parametric capacitance starts to shift towards negative detuning. Additionally, we can find $T$ from the FWHM at $g\mu_BB/\Delta_\text{c}>1$ and $g$ from the shift in detuning of the capacitance maximum as a function of $B$.

\subsubsection{Intermediate Relaxation Regime}

\begin{figure}[htbp]
	\centering
		\includegraphics{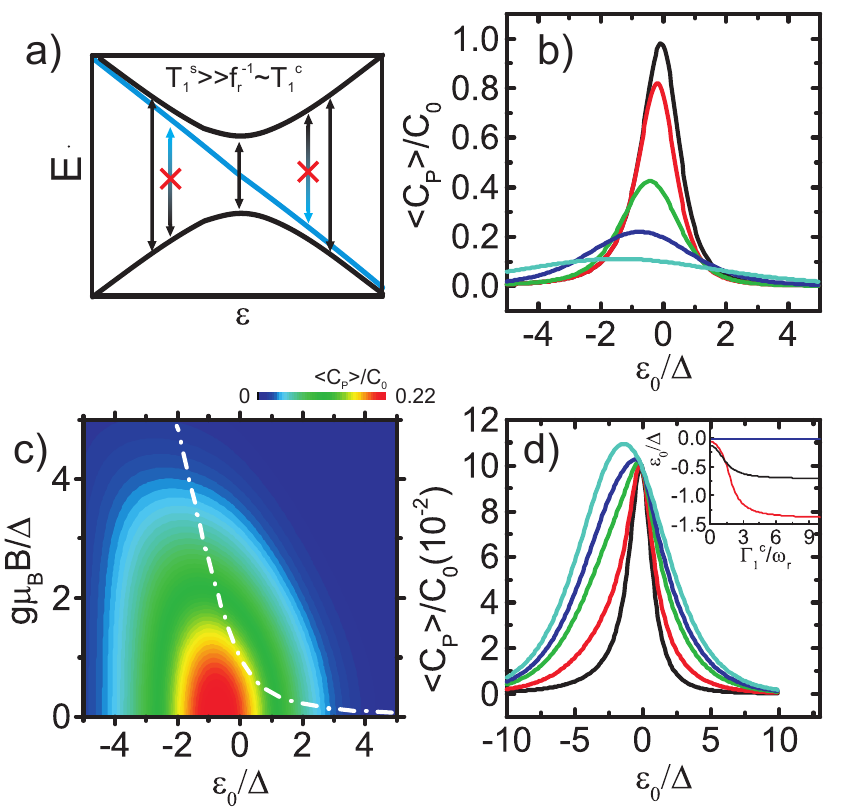}
	\caption{Spin qubit in the intermediate regime. (a) Schematic of the relaxation rates. Relaxation between states with same spin occurs within the time scale of the resonator whereas between states with different spin it does not. (b) Parametric capacitance lineshape as a function of detuning for $k_\text{c}T/\Delta_\text{c}=0.1,0.2,0.5,1,2$ (black, red, green, dark blue and light blue lines) and $\Gamma_1^\text{c}=10$. (c) Parametric capacitance as a function of detuning and magnetic field at $k_\text{c}T/\Delta_\text{c}=1$ and $\Gamma_1^\text{c}/\omega_\text{r}=10$. (d) Parametric capacitance lineshape as a function of detuning for $\Gamma_1^\text{c}/\omega_\text{r}=0,0.5,1,1.5,5$ (black, red, green, dark blue and light blue lines) and $k_\text{B}T/\Delta_\text{c}=1$. Inset: Maximum parametric capacitance position in detuning as a function of $\Gamma_1^\text{c}/\omega_\text{r}$ for $k_\text{B}T/\Delta_\text{c}=0.1, 1, 2$ (blue, black and red lines).}
	\label{fig:Fig5}
\end{figure}

Finally, we explore the singlet-triplet system in the intermediate relaxation limit $\Gamma_{1}^\text{s} \ll \omega_\text{r} \approx \Gamma_1^\text{c}$ using a similar time-dependent analysis as in section III, now including the triplet states. In this limit, the relaxation between states with different total spin numbers does not occur in the time scale of the resonator whereas relaxation between states with the same spin number does, as depicted in Fig.~\ref{fig:Fig5}(a). Under these conditions, the total singlet probability remains constant $\partial\chi_\text{s}^{'}/\partial\varepsilon=0$ and the third term in equation Eq.~\ref{eq:19} vanishes. In this limit, the parametric capacitance averaged over a resonator cycle can be written as


\begin{equation}\label{CQparaspin}
	\begin{aligned}
	&\left\langle C_\text{p}\right\rangle\approx C_0\left\{\frac{\Delta_\text{c}^3}{\Delta E^3(\varepsilon_0)}\chi_\text{s}^0\right. + \\
	&+\frac{2\Delta_\text{c}}{k_\text{B}T}\frac{\varepsilon_0^2}{\Delta E(\varepsilon_0)^2}f(\varepsilon_0,T,B)\frac{\Gamma_1^\text{c2}}{\omega_\text{r}^2+g^2(\varepsilon_0,T)\Gamma_1^\text{c2}}\biggl\},
	\end{aligned}
\end{equation}

where $f(\varepsilon_0,T,B)=\frac{\left(e^{-\Delta E(\varepsilon_0)/2k_\text{B}T}+e^{-3\Delta E(\varepsilon_0)/2k_\text{B}T}\right)}{Z_\text{s}^0}$.

In Fig.~\ref{fig:Fig5}(a) we plot the dependence of the parametric capacitance with detuning for various temperatures and $\Gamma_1^\text{c}/\omega_\text{r}=10$. We observe a decay in the peak height as in previous regimes and a shift of the maximum towards negative detunings following a linear dependence with temperature, $\left|\varepsilon_0\right|\approx 0.68k_\text{B}T$. This feature provides a simple measurement to identify the intermediate regime relaxation regime.

Additionally, in Fig.~\ref{fig:Fig5}(c), we plot the parametric capacitance lineshape as a function of increasing magnetic field for $\Gamma_1^\text{c}/\omega_\text{r}=10$ and $k_\text{B}T/\Delta_\text{c}=1$. We observe that the signal vanishes at high fields tracking the position of the singlet-triplet crossing point (dashed white line). This is in contrast to the slow relaxation regime in which, at this temperature, the signal vanished quasi-symmetrically, see Fig.\ref{fig:3}(d).

Finally, we investigate the dependence of the lineshape with increasing charge relaxation rate in Fig.~\ref{fig:Fig5}(c). We note two main features as a function of increasing $\Gamma_1^\text{c}$: An asymmetric increase in the width of the peak and a temperature-dependent shift of the position of the maximum towards negative detunings. The dependence with temperature can be seen more clearly in the inset where we plot the position in detuning of the maximum as a function of $\Gamma_1^\text{c}/\omega_\text{r}$. At $k_\text{B}T/\Delta_\text{c}=0.1$ (blue line) the maximum does not shift. However, as the temperature increases, the peak progressively shifts towards $\left|\varepsilon_0\right|\approx 0.68k_\text{B}T$ where it saturates at large $\Gamma_1^\text{c}$.

\section{V. CONCLUSION}

We have presented a theoretical derivation of the high-frequency capacitance of a DQD in the charge and spin qubit regimes. Our results demonstrate that the total capacitance of the system is composed by two terms with distinct physical origin: The quantum capacitance arising from adiabatic charge transitions and the finite curvature of the energy bands and the tunnelling capacitance that appears when population redistribution processes occur in the time-scale of the resonator period. Overall, our results provide a theoretical framework in which to understand the dependence of the qubit-resonator interaction on the qubit state and external parameters such as temperature and magnetic field. We obtain information about the charge $T_{1}^\text{c}$ and spin relaxation time $T_{1}^\text{s}$, tunnel coupling $\Delta_\text{c}$, electron temperature $T$ and electron g-factor $g$. Lastly, we note our analysis can be easily adapted to any quantum multi-level system such as solid-state qubit implementation like superconducting qubits~\cite{Pashkin2009} or hybrid qubits~\cite{Koh2012}. These results become of increasing interest for quantum state readout due to the benefits provided by in-situ gate-sensing - non-invasive, compact, high sensitivity and broadband detection - that open up a window for integrated qubit architectures.

\section{ACKNOWLEDGEMENTS}

We thank useful discussions with S. N. Shevchenko and D. A. Williams. This research is supported by the European Community's Seventh Framework Programme (FP7/2007-2013) through Grant Agreement No. 318397 (http://www.tolop.eu.) and Horizon 2020 through Grant Agreement 688539.



\begin{thebibliography}{34}
\expandafter\ifx\csname natexlab\endcsname\relax\def\natexlab#1{#1}\fi
\expandafter\ifx\csname bibnamefont\endcsname\relax
  \def\bibnamefont#1{#1}\fi
\expandafter\ifx\csname bibfnamefont\endcsname\relax
  \def\bibfnamefont#1{#1}\fi
\expandafter\ifx\csname citenamefont\endcsname\relax
  \def\citenamefont#1{#1}\fi
\expandafter\ifx\csname url\endcsname\relax
  \def\url#1{\texttt{#1}}\fi
\expandafter\ifx\csname urlprefix\endcsname\relax\def\urlprefix{URL }\fi
\providecommand{\bibinfo}[2]{#2}
\providecommand{\eprint}[2][]{\url{#2}}

\bibitem[{\citenamefont{Wallraff et~al.}(2004)\citenamefont{Wallraff, Schuster,
  Blais, Frunzio, Huang, Kumar, Girvin, and Schoelkopf}}]{Wallraff2004}
\bibinfo{author}{\bibfnamefont{A.}~\bibnamefont{Wallraff}},
  \bibinfo{author}{\bibfnamefont{D.~I.} \bibnamefont{Schuster}},
  \bibinfo{author}{\bibfnamefont{A.}~\bibnamefont{Blais}},
  \bibinfo{author}{\bibfnamefont{L.}~\bibnamefont{Frunzio}},
  \bibinfo{author}{\bibfnamefont{J.}~\bibnamefont{Huang},
  \bibfnamefont{R.~S.and~Majer}},
  \bibinfo{author}{\bibfnamefont{S.}~\bibnamefont{Kumar}},
  \bibinfo{author}{\bibfnamefont{S.~M.} \bibnamefont{Girvin}},
  \bibnamefont{and} \bibinfo{author}{\bibfnamefont{R.~J.}
  \bibnamefont{Schoelkopf}}, \bibinfo{journal}{Nature}
  \textbf{\bibinfo{volume}{431}}, \bibinfo{pages}{162} (\bibinfo{year}{2004}).

\bibitem[{\citenamefont{Petersson et~al.}(2012)\citenamefont{Petersson, McFaul,
  Schroer, Jung, Taylor, Houck, and Petta}}]{Petersson2012}
\bibinfo{author}{\bibfnamefont{K.~D.} \bibnamefont{Petersson}},
  \bibinfo{author}{\bibfnamefont{L.~W.} \bibnamefont{McFaul}},
  \bibinfo{author}{\bibfnamefont{M.~D.} \bibnamefont{Schroer}},
  \bibinfo{author}{\bibfnamefont{M.}~\bibnamefont{Jung}},
  \bibinfo{author}{\bibfnamefont{J.~M.} \bibnamefont{Taylor}},
  \bibinfo{author}{\bibfnamefont{A.~A.} \bibnamefont{Houck}}, \bibnamefont{and}
  \bibinfo{author}{\bibfnamefont{J.~R.} \bibnamefont{Petta}},
  \bibinfo{journal}{Nature} \textbf{\bibinfo{volume}{490}},
  \bibinfo{pages}{380} (\bibinfo{year}{2012}).

\bibitem[{\citenamefont{Colless et~al.}(2013)\citenamefont{Colless, Mahoney,
  Hornibrook, Doherty, Lu, Gossard, and Reilly}}]{Colless2013}
\bibinfo{author}{\bibfnamefont{J.~I.} \bibnamefont{Colless}},
  \bibinfo{author}{\bibfnamefont{A.~C.} \bibnamefont{Mahoney}},
  \bibinfo{author}{\bibfnamefont{J.~M.} \bibnamefont{Hornibrook}},
  \bibinfo{author}{\bibfnamefont{A.~C.} \bibnamefont{Doherty}},
  \bibinfo{author}{\bibfnamefont{H.}~\bibnamefont{Lu}},
  \bibinfo{author}{\bibfnamefont{A.~C.} \bibnamefont{Gossard}},
  \bibnamefont{and} \bibinfo{author}{\bibfnamefont{D.~J.}
  \bibnamefont{Reilly}}, \bibinfo{journal}{Physical Review Letters}
  \textbf{\bibinfo{volume}{110}}, \bibinfo{pages}{046805}
  (\bibinfo{year}{2013}).

\bibitem[{\citenamefont{Gonzalez-Zalba
  et~al.}(2015)\citenamefont{Gonzalez-Zalba, Barraud, Ferguson, and
  Betz}}]{Gonzalez-Zalba2015}
\bibinfo{author}{\bibfnamefont{M.~F.} \bibnamefont{Gonzalez-Zalba}},
  \bibinfo{author}{\bibfnamefont{S.}~\bibnamefont{Barraud}},
  \bibinfo{author}{\bibfnamefont{A.}~\bibnamefont{Ferguson}}, \bibnamefont{and}
  \bibinfo{author}{\bibfnamefont{A.~C.} \bibnamefont{Betz}},
  \bibinfo{journal}{Nat Commun} \textbf{\bibinfo{volume}{6}},
  \bibinfo{pages}{6084} (\bibinfo{year}{2015}).
	
\bibitem[{\citenamefont{Viennot
  et~al.}(2010)\citenamefont{Viennot, Dartiailh, Cottet, and Kontos}}]{Viennot2015}
\bibinfo{author}{\bibfnamefont{J.~J.} \bibnamefont{Viennot}},
  \bibinfo{author}{\bibfnamefont{M.~C.}~\bibnamefont{Dartiailh}},
	\bibinfo{author}{\bibfnamefont{A.}~\bibnamefont{Cottet}}, \bibnamefont{and}
  \bibinfo{author}{\bibfnamefont{T.} \bibnamefont{Kontos}},
  \bibinfo{journal}{Science} \textbf{\bibinfo{volume}{349}},
  \bibinfo{pages}{6246} (\bibinfo{year}{2015}).

\bibitem[{\citenamefont{Petersson et~al.}(2010)\citenamefont{Petersson, Smith,
  Anderson, Atkinson, Jones, and Ritchie}}]{Petersson2010}
\bibinfo{author}{\bibfnamefont{K.~D.} \bibnamefont{Petersson}},
  \bibinfo{author}{\bibfnamefont{C.~G.} \bibnamefont{Smith}},
  \bibinfo{author}{\bibfnamefont{D.}~\bibnamefont{Anderson}},
  \bibinfo{author}{\bibfnamefont{P.}~\bibnamefont{Atkinson}},
  \bibinfo{author}{\bibfnamefont{G.~A.~C.} \bibnamefont{Jones}},
  \bibnamefont{and} \bibinfo{author}{\bibfnamefont{D.~A.}
  \bibnamefont{Ritchie}}, \bibinfo{journal}{Nano Letters}
  \textbf{\bibinfo{volume}{10}}, \bibinfo{pages}{2789} (\bibinfo{year}{2010}).

\bibitem[{\citenamefont{Schoelkopf et~al.}(1998)\citenamefont{Schoelkopf,
  Wahlgren, Kozhevnikov, Delsing, and Prober}}]{Schoelkopf1998}
\bibinfo{author}{\bibfnamefont{R.~J.} \bibnamefont{Schoelkopf}},
  \bibinfo{author}{\bibfnamefont{P.}~\bibnamefont{Wahlgren}},
  \bibinfo{author}{\bibfnamefont{A.~A.} \bibnamefont{Kozhevnikov}},
  \bibinfo{author}{\bibfnamefont{P.}~\bibnamefont{Delsing}}, \bibnamefont{and}
  \bibinfo{author}{\bibfnamefont{D.~E.} \bibnamefont{Prober}},
  \bibinfo{journal}{Science} \textbf{\bibinfo{volume}{280}},
  \bibinfo{pages}{1238} (\bibinfo{year}{1998}).

\bibitem[{\citenamefont{Angus et~al.}(2008)\citenamefont{Angus, Ferguson,
  Dzurak, and Clark}}]{Angus2008}
\bibinfo{author}{\bibfnamefont{S.~J.} \bibnamefont{Angus}},
  \bibinfo{author}{\bibfnamefont{A.~J.} \bibnamefont{Ferguson}},
  \bibinfo{author}{\bibfnamefont{A.~S.} \bibnamefont{Dzurak}},
  \bibnamefont{and} \bibinfo{author}{\bibfnamefont{R.~G.} \bibnamefont{Clark}},
  \bibinfo{journal}{Applied Physics Letters} \textbf{\bibinfo{volume}{92}},
  \bibinfo{pages}{112103} (\bibinfo{year}{2008}).

\bibitem[{\citenamefont{Barthel et~al.}(2009)\citenamefont{Barthel, Reilly,
  Marcus, Hanson, and Gossard}}]{Barthel2009}
\bibinfo{author}{\bibfnamefont{C.}~\bibnamefont{Barthel}},
  \bibinfo{author}{\bibfnamefont{D.~J.} \bibnamefont{Reilly}},
  \bibinfo{author}{\bibfnamefont{C.~M.} \bibnamefont{Marcus}},
  \bibinfo{author}{\bibfnamefont{M.~P.} \bibnamefont{Hanson}},
  \bibnamefont{and} \bibinfo{author}{\bibfnamefont{A.~C.}
  \bibnamefont{Gossard}}, \bibinfo{journal}{Phys. Rev. Lett.}
  \textbf{\bibinfo{volume}{103}}, \bibinfo{pages}{160503}
  (\bibinfo{year}{2009}).

\bibitem[{\citenamefont{Chorley et~al.}(2012)\citenamefont{Chorley, Wabnig,
  Penfold-Fitch, Petersson, Frake, Smith, and Buitelaar}}]{Chorley2012}
\bibinfo{author}{\bibfnamefont{S.~J.} \bibnamefont{Chorley}},
  \bibinfo{author}{\bibfnamefont{J.}~\bibnamefont{Wabnig}},
  \bibinfo{author}{\bibfnamefont{Z.~V.} \bibnamefont{Penfold-Fitch}},
  \bibinfo{author}{\bibfnamefont{K.~D.} \bibnamefont{Petersson}},
  \bibinfo{author}{\bibfnamefont{J.}~\bibnamefont{Frake}},
  \bibinfo{author}{\bibfnamefont{C.~G.} \bibnamefont{Smith}}, \bibnamefont{and}
  \bibinfo{author}{\bibfnamefont{M.~R.} \bibnamefont{Buitelaar}},
  \bibinfo{journal}{Phys. Rev. Lett.} \textbf{\bibinfo{volume}{108}},
  \bibinfo{pages}{036802} (\bibinfo{year}{2012}).

\bibitem[{\citenamefont{Ciccarelli and Ferguson}(2011)}]{Ciccarelli2011}
\bibinfo{author}{\bibfnamefont{C.}~\bibnamefont{Ciccarelli}} \bibnamefont{and}
  \bibinfo{author}{\bibfnamefont{A.~J.} \bibnamefont{Ferguson}},
  \bibinfo{journal}{New J. Phys.} \textbf{\bibinfo{volume}{13}},
  \bibinfo{pages}{093015} (\bibinfo{year}{2011}).

\bibitem[{\citenamefont{Cottet et~al.}(2011)\citenamefont{Cottet, Mora, and
  Kontos}}]{Cottet2011}
\bibinfo{author}{\bibfnamefont{A.}~\bibnamefont{Cottet}},
  \bibinfo{author}{\bibfnamefont{C.}~\bibnamefont{Mora}}, \bibnamefont{and}
  \bibinfo{author}{\bibfnamefont{T.}~\bibnamefont{Kontos}},
  \bibinfo{journal}{Phys. Rev. B} \textbf{\bibinfo{volume}{83}},
  \bibinfo{pages}{121311} (\bibinfo{year}{2011}).

\bibitem[{\citenamefont{Schroer et~al.}(2012)\citenamefont{Schroer, Jung,
  Petersson, and Petta}}]{Schroer2012}
\bibinfo{author}{\bibfnamefont{M.~D.} \bibnamefont{Schroer}},
  \bibinfo{author}{\bibfnamefont{M.}~\bibnamefont{Jung}},
  \bibinfo{author}{\bibfnamefont{K.~D.} \bibnamefont{Petersson}},
  \bibnamefont{and} \bibinfo{author}{\bibfnamefont{J.~R.} \bibnamefont{Petta}},
  \bibinfo{journal}{Phys. Rev. Lett.} \textbf{\bibinfo{volume}{109}},
  \bibinfo{pages}{166804} (\bibinfo{year}{2012}).

\bibitem[{\citenamefont{Sillanp\"a\"a et~al.}(2005)\citenamefont{Sillanp\"a\"a,
  Lehtinen, Paila, Makhlin, Roschier, and Hakonen}}]{Sillanpaa2005}
\bibinfo{author}{\bibfnamefont{M.~A.} \bibnamefont{Sillanp\"a\"a}},
  \bibinfo{author}{\bibfnamefont{T.}~\bibnamefont{Lehtinen}},
  \bibinfo{author}{\bibfnamefont{A.}~\bibnamefont{Paila}},
  \bibinfo{author}{\bibfnamefont{Y.}~\bibnamefont{Makhlin}},
  \bibinfo{author}{\bibfnamefont{L.}~\bibnamefont{Roschier}}, \bibnamefont{and}
  \bibinfo{author}{\bibfnamefont{P.~J.} \bibnamefont{Hakonen}},
  \bibinfo{journal}{Phys. Rev. Lett.} \textbf{\bibinfo{volume}{95}},
  \bibinfo{pages}{206806} (\bibinfo{year}{2005}).

\bibitem[{\citenamefont{Duty et~al.}(2005)\citenamefont{Duty, Johansson, Bladh,
  Gunnarsson, Wilson, and Delsing}}]{Duty2005}
\bibinfo{author}{\bibfnamefont{T.}~\bibnamefont{Duty}},
  \bibinfo{author}{\bibfnamefont{G.}~\bibnamefont{Johansson}},
  \bibinfo{author}{\bibfnamefont{K.}~\bibnamefont{Bladh}},
  \bibinfo{author}{\bibfnamefont{D.}~\bibnamefont{Gunnarsson}},
  \bibinfo{author}{\bibfnamefont{C.}~\bibnamefont{Wilson}}, \bibnamefont{and}
  \bibinfo{author}{\bibfnamefont{P.}~\bibnamefont{Delsing}},
  \bibinfo{journal}{Phys. Rev. Lett.} \textbf{\bibinfo{volume}{95}},
  \bibinfo{pages}{206807} (\bibinfo{year}{2005}).

\bibitem[{\citenamefont{Persson
  et~al.}(2010{\natexlab{a}})\citenamefont{Persson, Wilson, Sandberg, and
  Delsing}}]{Persson2010}
\bibinfo{author}{\bibfnamefont{F.}~\bibnamefont{Persson}},
  \bibinfo{author}{\bibfnamefont{C.~M.} \bibnamefont{Wilson}},
  \bibinfo{author}{\bibfnamefont{M.}~\bibnamefont{Sandberg}}, \bibnamefont{and}
  \bibinfo{author}{\bibfnamefont{P.}~\bibnamefont{Delsing}},
  \bibinfo{journal}{Phys. Rev. B} \textbf{\bibinfo{volume}{82}},
  \bibinfo{pages}{134533} (\bibinfo{year}{2010}{\natexlab{a}}).

\bibitem[{\citenamefont{Betz et~al.}(2015)\citenamefont{Betz, Wacquez, Vinet,
  Jehl, Saraiva, Sanquer, Ferguson, and Gonzalez-Zalba}}]{Betz2015}
\bibinfo{author}{\bibfnamefont{A.~C.} \bibnamefont{Betz}},
  \bibinfo{author}{\bibfnamefont{R.}~\bibnamefont{Wacquez}},
  \bibinfo{author}{\bibfnamefont{M.}~\bibnamefont{Vinet}},
  \bibinfo{author}{\bibfnamefont{X.}~\bibnamefont{Jehl}},
  \bibinfo{author}{\bibfnamefont{A.~L.} \bibnamefont{Saraiva}},
  \bibinfo{author}{\bibfnamefont{M.}~\bibnamefont{Sanquer}},
  \bibinfo{author}{\bibfnamefont{A.~J.} \bibnamefont{Ferguson}},
  \bibnamefont{and} \bibinfo{author}{\bibfnamefont{M.~F.}
  \bibnamefont{Gonzalez-Zalba}}, \bibinfo{journal}{Nano Letters}
  \textbf{\bibinfo{volume}{15}}, \bibinfo{pages}{4622} (\bibinfo{year}{2015}).

\bibitem[{\citenamefont{Ashoori et~al.}(1992)\citenamefont{Ashoori, Stormer,
  Weiner, Pfeiffer, Pearton, Baldwin, and West}}]{Ashoori1992}
\bibinfo{author}{\bibfnamefont{R.~C.} \bibnamefont{Ashoori}},
  \bibinfo{author}{\bibfnamefont{H.~L.} \bibnamefont{Stormer}},
  \bibinfo{author}{\bibfnamefont{J.~S.} \bibnamefont{Weiner}},
  \bibinfo{author}{\bibfnamefont{L.~N.} \bibnamefont{Pfeiffer}},
  \bibinfo{author}{\bibfnamefont{S.~J.} \bibnamefont{Pearton}},
  \bibinfo{author}{\bibfnamefont{K.~W.} \bibnamefont{Baldwin}},
  \bibnamefont{and} \bibinfo{author}{\bibfnamefont{K.~W.} \bibnamefont{West}},
  \bibinfo{journal}{Phys. Rev. Lett.} \textbf{\bibinfo{volume}{68}},
  \bibinfo{pages}{3088} (\bibinfo{year}{1992}).

\bibitem[{\citenamefont{Persson
  et~al.}(2010{\natexlab{b}})\citenamefont{Persson, Wilson, Sandberg,
  Johansson, and Delsing}}]{Persson2010a}
\bibinfo{author}{\bibfnamefont{F.}~\bibnamefont{Persson}},
  \bibinfo{author}{\bibfnamefont{C.~M.} \bibnamefont{Wilson}},
  \bibinfo{author}{\bibfnamefont{M.}~\bibnamefont{Sandberg}},
  \bibinfo{author}{\bibfnamefont{G.}~\bibnamefont{Johansson}},
  \bibnamefont{and} \bibinfo{author}{\bibfnamefont{P.}~\bibnamefont{Delsing}},
  \bibinfo{journal}{Nano Letters} \textbf{\bibinfo{volume}{10}},
  \bibinfo{pages}{953} (\bibinfo{year}{2010}{\natexlab{b}}).

\bibitem[{\citenamefont{Gonzalez-Zalba et~al.}(0)\citenamefont{Gonzalez-Zalba,
  Shevchenko, Barraud, Johansson, Ferguson, Nori, and
  Betz}}]{Gonzalez-Zalba2016}
\bibinfo{author}{\bibfnamefont{M.~F.} \bibnamefont{Gonzalez-Zalba}},
  \bibinfo{author}{\bibfnamefont{S.~N.} \bibnamefont{Shevchenko}},
  \bibinfo{author}{\bibfnamefont{S.}~\bibnamefont{Barraud}},
  \bibinfo{author}{\bibfnamefont{J.~R.} \bibnamefont{Johansson}},
  \bibinfo{author}{\bibfnamefont{A.~J.} \bibnamefont{Ferguson}},
  \bibinfo{author}{\bibfnamefont{F.}~\bibnamefont{Nori}}, \bibnamefont{and}
  \bibinfo{author}{\bibfnamefont{A.~C.} \bibnamefont{Betz}},
  \bibinfo{journal}{Nano Letters} \textbf{\bibinfo{volume}{0}},
  \bibinfo{pages}{null} (\bibinfo{year}{0}).

\bibitem[{\citenamefont{Kim et~al.}(2015)\citenamefont{Kim, Ward, Simmons,
  Gamble, Blume-Kohout, Nielsen, Savage, Lagally, Friesen, Coppersmith
  et~al.}}]{Kim2015}
\bibinfo{author}{\bibfnamefont{D.}~\bibnamefont{Kim}},
  \bibinfo{author}{\bibfnamefont{D.~R.} \bibnamefont{Ward}},
  \bibinfo{author}{\bibfnamefont{C.~B.} \bibnamefont{Simmons}},
  \bibinfo{author}{\bibfnamefont{J.~K.} \bibnamefont{Gamble}},
  \bibinfo{author}{\bibfnamefont{R.}~\bibnamefont{Blume-Kohout}},
  \bibinfo{author}{\bibfnamefont{E.}~\bibnamefont{Nielsen}},
  \bibinfo{author}{\bibfnamefont{D.~E.} \bibnamefont{Savage}},
  \bibinfo{author}{\bibfnamefont{M.~G.} \bibnamefont{Lagally}},
  \bibinfo{author}{\bibfnamefont{M.}~\bibnamefont{Friesen}},
  \bibinfo{author}{\bibfnamefont{S.~N.} \bibnamefont{Coppersmith}},
  \bibnamefont{et~al.}, \bibinfo{journal}{Nat. Nano.}
  \textbf{\bibinfo{volume}{10}}, \bibinfo{pages}{243} (\bibinfo{year}{2015}).

\bibitem[{\citenamefont{Veldhorst et~al.}(2015)\citenamefont{Veldhorst, Yang,
  Hwang, Huang, Dehollain, Muhonen, Simmons, Laucht, Hudson, Itoh
  et~al.}}]{Veldhorst2015}
\bibinfo{author}{\bibfnamefont{M.}~\bibnamefont{Veldhorst}},
  \bibinfo{author}{\bibfnamefont{C.~H.} \bibnamefont{Yang}},
  \bibinfo{author}{\bibfnamefont{J.~C.~C.} \bibnamefont{Hwang}},
  \bibinfo{author}{\bibfnamefont{W.}~\bibnamefont{Huang}},
  \bibinfo{author}{\bibfnamefont{J.~P.} \bibnamefont{Dehollain}},
  \bibinfo{author}{\bibfnamefont{J.~T.} \bibnamefont{Muhonen}},
  \bibinfo{author}{\bibfnamefont{S.}~\bibnamefont{Simmons}},
  \bibinfo{author}{\bibfnamefont{A.}~\bibnamefont{Laucht}},
  \bibinfo{author}{\bibfnamefont{F.~E.} \bibnamefont{Hudson}},
  \bibinfo{author}{\bibfnamefont{K.~M.} \bibnamefont{Itoh}},
  \bibnamefont{et~al.}, \bibinfo{journal}{Nature}
  \textbf{\bibinfo{volume}{526}}, \bibinfo{pages}{410} (\bibinfo{year}{2015}).

\bibitem[{\citenamefont{Pashkin et~al.}(2009)\citenamefont{Pashkin, Astafiev,
  Yamamoto, Nakamura, and Tsai}}]{Pashkin2009}
\bibinfo{author}{\bibfnamefont{Y.~A.} \bibnamefont{Pashkin}},
  \bibinfo{author}{\bibfnamefont{O.}~\bibnamefont{Astafiev}},
  \bibinfo{author}{\bibfnamefont{T.}~\bibnamefont{Yamamoto}},
  \bibinfo{author}{\bibfnamefont{Y.}~\bibnamefont{Nakamura}}, \bibnamefont{and}
  \bibinfo{author}{\bibfnamefont{J.~S.} \bibnamefont{Tsai}},
  \bibinfo{journal}{Quantum Information Processing}
  \textbf{\bibinfo{volume}{8}}, \bibinfo{pages}{55} (\bibinfo{year}{2009}).

\bibitem[{\citenamefont{Koh et~al.}(2012)\citenamefont{Koh, Gamble, Friesen,
  Eriksson, and Coppersmith}}]{Koh2012}
\bibinfo{author}{\bibfnamefont{T.~S.} \bibnamefont{Koh}},
  \bibinfo{author}{\bibfnamefont{J.~K.} \bibnamefont{Gamble}},
  \bibinfo{author}{\bibfnamefont{M.}~\bibnamefont{Friesen}},
  \bibinfo{author}{\bibfnamefont{M.~A.} \bibnamefont{Eriksson}},
  \bibnamefont{and} \bibinfo{author}{\bibfnamefont{S.~N.}
  \bibnamefont{Coppersmith}}, \bibinfo{journal}{Phys. Rev. Lett.}
  \textbf{\bibinfo{volume}{109}}, \bibinfo{pages}{250503}
  (\bibinfo{year}{2012}).
	
\bibitem[{\citenamefont{Shevchenko et~al.}(2012)\citenamefont{Shevchenko, Omelyanchouk, and Il'ichev}}]{Shevchenko2012}
\bibinfo{author}{\bibfnamefont{S.~N.} \bibnamefont{Shevchenko}},
  \bibinfo{author}{\bibfnamefont{A.~N.}~\bibnamefont{Omelyanchouk}},
  \bibnamefont{and} \bibinfo{author}{\bibfnamefont{E.}
  \bibnamefont{Il'ichev}}, \bibinfo{journal}{Low Temp. Phys.}
  \textbf{\bibinfo{volume}{38}}, \bibinfo{pages}{283} (\bibinfo{year}{2012}).

\bibitem[{\citenamefont{LeHaye et~al.}(2009)\citenamefont{LeHaye, Suh,
  Echternach, Schwab, and Roukes}}]{LeHaye2009}
\bibinfo{author}{\bibfnamefont{M.~D.} \bibnamefont{LeHaye}},
  \bibinfo{author}{\bibfnamefont{J.}~\bibnamefont{Suh}},
  \bibinfo{author}{\bibfnamefont{P.~M.} \bibnamefont{Echternach}},
  \bibinfo{author}{\bibfnamefont{K.~C.} \bibnamefont{Schwab}},
  \bibnamefont{and} \bibinfo{author}{\bibfnamefont{M.~L.}
  \bibnamefont{Roukes}}, \bibinfo{journal}{Nature}
  \textbf{\bibinfo{volume}{459}}, \bibinfo{pages}{960} (\bibinfo{year}{2009}).

\bibitem[{\citenamefont{Shevchenko et~al.}(2012)\citenamefont{Shevchenko, Ashhab, and Nori}}]{Shevchenko2012b}
\bibinfo{author}{\bibfnamefont{S.~N.} \bibnamefont{Shevchenko}},
  \bibinfo{author}{\bibfnamefont{S.}~\bibnamefont{Ashhab}},
  \bibnamefont{and} \bibinfo{author}{\bibfnamefont{F.}
  \bibnamefont{Nori}}, \bibinfo{journal}{Phys. Rev. B}
  \textbf{\bibinfo{volume}{85}}, \bibinfo{pages}{094502} (\bibinfo{year}{2012}).

\bibitem[{\citenamefont{Gonzalez-Zalba
  et~al.}(2014)\citenamefont{Gonzalez-Zalba, Saraiva, Calderon, Heiss, Koiller,
  and Ferguson}}]{Gonzalez-Zalba2014}
\bibinfo{author}{\bibfnamefont{M.~F.} \bibnamefont{Gonzalez-Zalba}},
  \bibinfo{author}{\bibfnamefont{A.}~\bibnamefont{Saraiva}},
  \bibinfo{author}{\bibfnamefont{M.~J.} \bibnamefont{Calderon}},
  \bibinfo{author}{\bibfnamefont{D.}~\bibnamefont{Heiss}},
  \bibinfo{author}{\bibfnamefont{B.}~\bibnamefont{Koiller}}, \bibnamefont{and}
  \bibinfo{author}{\bibfnamefont{A.~J.} \bibnamefont{Ferguson}},
  \bibinfo{journal}{Nano Letters} \textbf{\bibinfo{volume}{14}},
  \bibinfo{pages}{5672} (\bibinfo{year}{2014}).

\bibitem[{\citenamefont{Hile et~al.}(2015)\citenamefont{Hile, House, Peretz,
  Verduijn, Widmann, Kobayashi, Rogge, and Simmons}}]{Hile2015}
\bibinfo{author}{\bibfnamefont{S.~J.} \bibnamefont{Hile}},
  \bibinfo{author}{\bibfnamefont{M.~G.} \bibnamefont{House}},
  \bibinfo{author}{\bibfnamefont{E.}~\bibnamefont{Peretz}},
  \bibinfo{author}{\bibfnamefont{J.}~\bibnamefont{Verduijn}},
  \bibinfo{author}{\bibfnamefont{D.}~\bibnamefont{Widmann}},
  \bibinfo{author}{\bibfnamefont{T.}~\bibnamefont{Kobayashi}},
  \bibinfo{author}{\bibfnamefont{S.}~\bibnamefont{Rogge}}, \bibnamefont{and}
  \bibinfo{author}{\bibfnamefont{M.~Y.} \bibnamefont{Simmons}},
  \bibinfo{journal}{App. Phys. Lett.} \textbf{\bibinfo{volume}{107}}
  (\bibinfo{year}{2015}).

\bibitem[{\citenamefont{Frey et~al.}(2012)\citenamefont{Frey, Leek, Beck,
  Faist, Wallraff, Ensslin, Ihn, and B\"uttiker}}]{Frey2012}
\bibinfo{author}{\bibfnamefont{T.}~\bibnamefont{Frey}},
  \bibinfo{author}{\bibfnamefont{P.~J.} \bibnamefont{Leek}},
  \bibinfo{author}{\bibfnamefont{M.}~\bibnamefont{Beck}},
  \bibinfo{author}{\bibfnamefont{J.}~\bibnamefont{Faist}},
  \bibinfo{author}{\bibfnamefont{A.}~\bibnamefont{Wallraff}},
  \bibinfo{author}{\bibfnamefont{K.}~\bibnamefont{Ensslin}},
  \bibinfo{author}{\bibfnamefont{T.}~\bibnamefont{Ihn}}, \bibnamefont{and}
  \bibinfo{author}{\bibfnamefont{M.}~\bibnamefont{B\"uttiker}},
  \bibinfo{journal}{Phys. Rev. B} \textbf{\bibinfo{volume}{86}},
  \bibinfo{pages}{115303} (\bibinfo{year}{2012}).
	
\bibitem[{\citenamefont{Berns et~al.}(2006)\citenamefont{Berns, Oliver, Valenzuela,
  Shytov, Berggren, Levitov, and Orlando}}]{Berns2006}
\bibinfo{author}{\bibfnamefont{D.~M.}~\bibnamefont{Berns}},
  \bibinfo{author}{\bibfnamefont{W.~D.} \bibnamefont{Oliver}},
  \bibinfo{author}{\bibfnamefont{S.~O.}~\bibnamefont{Valenzuela}},
  \bibinfo{author}{\bibfnamefont{A.~V.}~\bibnamefont{Shytov}},
  \bibinfo{author}{\bibfnamefont{K.~K.}~\bibnamefont{Berggren}},
  \bibinfo{author}{\bibfnamefont{L.~S.}~\bibnamefont{Levitov}},
  \bibnamefont{and}
  \bibinfo{author}{\bibfnamefont{T.~P.}~\bibnamefont{Orlando}},
  \bibinfo{journal}{Phys. Rev. Lett.} \textbf{\bibinfo{volume}{97}},
  \bibinfo{pages}{150502} (\bibinfo{year}{2006}).

\bibitem[{\citenamefont{Petta et~al.}(2010)\citenamefont{Petta, Lu, and
  Gossard}}]{Petta2010}
\bibinfo{author}{\bibfnamefont{J.~R.} \bibnamefont{Petta}},
  \bibinfo{author}{\bibfnamefont{H.}~\bibnamefont{Lu}}, \bibnamefont{and}
  \bibinfo{author}{\bibfnamefont{A.~C.} \bibnamefont{Gossard}},
  \bibinfo{journal}{Science} \textbf{\bibinfo{volume}{327}},
  \bibinfo{pages}{669} (\bibinfo{year}{2010}).

\bibitem[{\citenamefont{Maune et~al.}(2012)\citenamefont{Maune, Borselli,
  Huang, Ladd, Deelman, Holabird, Kiselev, Alvarado-Rodriguez, Ross, Schmitz
  et~al.}}]{Maune2012}
\bibinfo{author}{\bibfnamefont{B.~M.} \bibnamefont{Maune}},
  \bibinfo{author}{\bibfnamefont{M.~G.} \bibnamefont{Borselli}},
  \bibinfo{author}{\bibfnamefont{B.}~\bibnamefont{Huang}},
  \bibinfo{author}{\bibfnamefont{T.~D.} \bibnamefont{Ladd}},
  \bibinfo{author}{\bibfnamefont{P.~W.} \bibnamefont{Deelman}},
  \bibinfo{author}{\bibfnamefont{K.~S.} \bibnamefont{Holabird}},
  \bibinfo{author}{\bibfnamefont{A.~A.} \bibnamefont{Kiselev}},
  \bibinfo{author}{\bibfnamefont{I.}~\bibnamefont{Alvarado-Rodriguez}},
  \bibinfo{author}{\bibfnamefont{R.~S.} \bibnamefont{Ross}},
  \bibinfo{author}{\bibfnamefont{A.~E.} \bibnamefont{Schmitz}},
  \bibnamefont{et~al.}, \bibinfo{journal}{Nature}
  \textbf{\bibinfo{volume}{481}}, \bibinfo{pages}{344} (\bibinfo{year}{2012}).

\bibitem[{\citenamefont{Lim et~al.}(2011)\citenamefont{Lim, Yang, Zwanenburg,
  and Dzurak}}]{Lim2011}
\bibinfo{author}{\bibfnamefont{W.~H.} \bibnamefont{Lim}},
  \bibinfo{author}{\bibfnamefont{C.~H.} \bibnamefont{Yang}},
  \bibinfo{author}{\bibfnamefont{F.~A.} \bibnamefont{Zwanenburg}},
  \bibnamefont{and} \bibinfo{author}{\bibfnamefont{A.~S.}
  \bibnamefont{Dzurak}}, \bibinfo{journal}{Nanotechnology}
  \textbf{\bibinfo{volume}{22}}, \bibinfo{pages}{335704}
  (\bibinfo{year}{2011}), ISSN \bibinfo{issn}{0957-4484}.

\bibitem[{\citenamefont{Petta et~al.}(2005)\citenamefont{Petta, Johnson,
  Taylor, Laird, Yacoby, Lukin, Marcus, Hanson, and Gossard}}]{Petta2005}
\bibinfo{author}{\bibfnamefont{J.~R.} \bibnamefont{Petta}},
  \bibinfo{author}{\bibfnamefont{A.~C.} \bibnamefont{Johnson}},
  \bibinfo{author}{\bibfnamefont{J.~M.} \bibnamefont{Taylor}},
  \bibinfo{author}{\bibfnamefont{E.~A.} \bibnamefont{Laird}},
  \bibinfo{author}{\bibfnamefont{A.}~\bibnamefont{Yacoby}},
  \bibinfo{author}{\bibfnamefont{M.~D.} \bibnamefont{Lukin}},
  \bibinfo{author}{\bibfnamefont{C.~M.} \bibnamefont{Marcus}},
  \bibinfo{author}{\bibfnamefont{M.~P.} \bibnamefont{Hanson}},
  \bibnamefont{and} \bibinfo{author}{\bibfnamefont{A.~C.}
  \bibnamefont{Gossard}}, \bibinfo{journal}{Science}
  \textbf{\bibinfo{volume}{309}}, \bibinfo{pages}{2180} (\bibinfo{year}{2005}).

\bibitem[{\citenamefont{Wu et~al.}(2014)\citenamefont{Wu, Ward, Prance, Kim,
  Gamble, Mohr, Shi, Savage, Lagally, Friesen et~al.}}]{Wu2014}
\bibinfo{author}{\bibfnamefont{X.}~\bibnamefont{Wu}},
  \bibinfo{author}{\bibfnamefont{D.~R.} \bibnamefont{Ward}},
  \bibinfo{author}{\bibfnamefont{J.~R.} \bibnamefont{Prance}},
  \bibinfo{author}{\bibfnamefont{D.}~\bibnamefont{Kim}},
  \bibinfo{author}{\bibfnamefont{J.~K.} \bibnamefont{Gamble}},
  \bibinfo{author}{\bibfnamefont{R.~T.} \bibnamefont{Mohr}},
  \bibinfo{author}{\bibfnamefont{Z.}~\bibnamefont{Shi}},
  \bibinfo{author}{\bibfnamefont{D.~E.} \bibnamefont{Savage}},
  \bibinfo{author}{\bibfnamefont{M.~G.} \bibnamefont{Lagally}},
  \bibinfo{author}{\bibfnamefont{M.}~\bibnamefont{Friesen}},
  \bibnamefont{et~al.}, \bibinfo{journal}{PNAS} \textbf{\bibinfo{volume}{111}},
  \bibinfo{pages}{11938} (\bibinfo{year}{2014}).

\bibitem[{\citenamefont{Urdampilleta et~al.}(2015)\citenamefont{Urdampilleta,
  Chatterjee, Lo, Kobayashi, Mansir, Barraud, Betz, Rogge, Gonzalez-Zalba, and
  Morton}}]{Urdampilleta2015}
\bibinfo{author}{\bibfnamefont{M.}~\bibnamefont{Urdampilleta}},
  \bibinfo{author}{\bibfnamefont{A.}~\bibnamefont{Chatterjee}},
  \bibinfo{author}{\bibfnamefont{C.~C.} \bibnamefont{Lo}},
  \bibinfo{author}{\bibfnamefont{T.}~\bibnamefont{Kobayashi}},
  \bibinfo{author}{\bibfnamefont{J.}~\bibnamefont{Mansir}},
  \bibinfo{author}{\bibfnamefont{S.}~\bibnamefont{Barraud}},
  \bibinfo{author}{\bibfnamefont{A.~C.} \bibnamefont{Betz}},
  \bibinfo{author}{\bibfnamefont{S.}~\bibnamefont{Rogge}},
  \bibinfo{author}{\bibfnamefont{M.~F.} \bibnamefont{Gonzalez-Zalba}},
  \bibnamefont{and} \bibinfo{author}{\bibfnamefont{J.~J.~L.}
  \bibnamefont{Morton}}, \bibinfo{journal}{Phys. Rev. X}
  \textbf{\bibinfo{volume}{5}}, \bibinfo{pages}{031024} (\bibinfo{year}{2015}).

\bibitem[{\citenamefont{House et~al.}(2015)\citenamefont{House, Kobayashi,
  Weber, Hile, Watson, van~der Heijden, Rogge, and Simmons}}]{House2015}
\bibinfo{author}{\bibfnamefont{M.~G.} \bibnamefont{House}},
  \bibinfo{author}{\bibfnamefont{K.}~\bibnamefont{Kobayashi}},
  \bibinfo{author}{\bibfnamefont{B.}~\bibnamefont{Weber}},
  \bibinfo{author}{\bibfnamefont{S.~J.} \bibnamefont{Hile}},
  \bibinfo{author}{\bibfnamefont{T.~F.} \bibnamefont{Watson}},
  \bibinfo{author}{\bibfnamefont{J.}~\bibnamefont{van~der Heijden}},
  \bibinfo{author}{\bibfnamefont{S.}~\bibnamefont{Rogge}}, \bibnamefont{and}
  \bibinfo{author}{\bibfnamefont{M.~Y.} \bibnamefont{Simmons}},
  \bibinfo{journal}{Nat Commun} \textbf{\bibinfo{volume}{6}},
  \bibinfo{pages}{8848} (\bibinfo{year}{2015}).

\end{thebibliography}


\end{document}